\documentclass[fleqn,usenatbib]{mnras}

\usepackage[T1]{fontenc}
\usepackage{ae,aecompl}
\usepackage{courier}

\usepackage{graphicx}	
\usepackage{amssymb}	
\graphicspath{{../images/}}
\usepackage{booktabs,array,caption}
\usepackage{epsfig}

\makeatletter 
\def\eqalign#1{\null\,\vcenter{\openup\jot\m@th  \ialign{\strut\hfil$\displaystyle{##}$&$\displaystyle{{}##}$\hfil      \crcr#1\crcr}}\,} 
\makeatother

\DeclareMathAlphabet{\mathsc}{OT1}{cmr}{m}{sc}
\def\testbx{bx}%
\DeclareRobustCommand{\ion}[2]{%
\relax\ifmmode
\ifx\testbx\f@series
{\mathbf{#1\,\mathsc{#2}}}\else
{\mathrm{#1\,\mathsc{#2}}}\fi
\else\textup{#1\,{\mdseries\textsc{#2}}}%
\fi}

\title[Effects of density on the oxygen ionisation equilibrium]{Effects of density on the oxygen ionisation equilibrium in collisional plasmas\thanks{Electron impact ionisation rate coefficients and ion populations are available at CDS via anonymous ftp to cdsarc.u-strasbg.fr (130.79.128.5) or via http://cdsarc.u-strasbg.fr/viz-bin/qcat?J/MNRAS.}}

\author[R.P. Dufresne et al.]{
R.P. Dufresne,$^{1}$\thanks{E-mail: rpd21@cam.ac.uk}
G. Del Zanna,$^{1}$
and N.R. Badnell$^{2}$
\\
$^{1}$DAMTP, University of Cambridge, Wilberforce Road, Cambridge CB3 0WA, UK\\
$^{2}$Department of Physics, University of Strathclyde, Glasgow G4 0NG, UK
}

\date{Accepted XXX. Received YYY; in original form ZZZ}

\pubyear{2020}

\begin{document}
\label{firstpage}
\pagerange{\pageref{firstpage}--\pageref{lastpage}}
\maketitle

\begin{abstract}
The ion populations most frequently adopted for diagnostics in collisional plasmas are derived from the density independent, coronal approximation. In higher density, lower temperature conditions, ionisation rates are enhanced once metastable levels become populated, and recombination rates are suppressed if ions recombine into Rydberg levels. As a result, the formation temperatures of ions shift, altering the diagnostics of the plasma. To accurately model the effect of ionisation from metastable levels, new electron impact, ionisation cross sections have been calculated for oxygen, both for direct ionisation and excitation--auto-ionisation of the ground and metastable levels. The results have been incorporated into collisional radiative modelling to show how the ionisation equilibrium of oxygen changes once metastable levels become populated. Suppression of dielectronic recombination has been estimated and also included in the modelling, demonstrating the shifts with density in comparison to the coronal approximation. The final results for the ionisation equilibrium are used in differential emission measure modelling to predict line intensities for many lines emitted by \ion{O}{ii}-\ion{O}{vi} in the solar transition region. The predictions show improved agreement by 15-40\% for \ion{O}{ii}, \ion{O}{vi} and the inter-combination lines of \ion{O}{iii}-\ion{O}{v}, when compared to results from coronal approximation modelling. While there are still discrepancies with observations of these lines, this could, to a large part, be explained by variability in the observations.
\end{abstract}




\section{Introduction}
\label{sec:intro}

With the advent of Solar Orbiter and Parker Solar Probe comes the capability of coupling in situ and remote sensing data, an opportunity which will reveal the link between the solar atmosphere and the wind which escapes from the Sun. A key component in the remote sensing observations of Solar Orbiter is the Spectral Imaging of the Coronal Environment (SPICE) instrument \citep{anderson2019}. A notable feature of SPICE is that it will observe lines from each of the first six charge states of oxygen. Simultaneous observations of these ions will make it possible to diagnose properties such as time dependent ionisation and the temperature distribution of the atmosphere over a wide range. An important component for these diagnostics is the accuracy of the oxygen ion populations at any given density and temperature. The method most commonly adopted for predicting charge state distributions is the coronal approximation, which is density independent. While this method is suitable for conditions present in the solar corona, where the density is lower than other parts of the atmosphere, it shows discrepancies when modelling TR lines, which are the nature of the oxygen lines observed by SPICE. 

An analysis by \cite{doschek1999} tests a number of lines from \ion{O}{iii}--\ion{O}{v} and finds that the observed and predicted line intensities for lines emitted by \ion{O}{iv} and \ion{O}{v}, when expressed as ratios with inter-combination lines, differ by a factor of two. They suggest many possibilities for these anomalies, such as variations in solar emission, inadequate atomic data and assuming the lines are emitted at a single temperature. The authors did not question directly, however, the accuracy of the ion populations from \cite{arnaud1985}, calculated in the coronal approximation, which they used to predict the line intensities. 

The discrepancy between modelling and observations caused by the coronal approximation is most apparent with Li- and Na-like TR ions, for which observed line intensities are factors of approximately two to eight times stronger than intensities predicted by the modelling. The primary causes of the differences arise from: suppression of dielectronic recombination (DR), first shown by \cite{burgess1969}, and the influence of metastable levels on the ion balance, which \cite{summers1983} demonstrate. \cite{young2018} modifies the coronal approximation with an estimate of DR suppression, and finds that the ratio of the neon to oxygen abundances alters as a result. \cite{doyle2005} used the more sophisticated atomic modelling of the Atomic Data and Analysis Structure (ADAS), which includes both of the density effects mentioned, to show how predicted intensities of lines from Li-like ions formed in the TR would increase by 15-50\% for the higher temperature lines and by up to three times for \ion{C}{iv}. 

\cite{dufresne2019} modify the coronal approximation by including metastable levels and an estimation of the suppression of DR, and find not only an improvement for \ion{C}{iv}, but that predicted line intensities come into closer agreement with observations for all the charge states of carbon which form in the solar TR. Consequently, the aim here is to extend the same modelling to oxygen, in order to provide ion populations which will be suitable for use with data from Solar Orbiter and similar missions. They will also be appropriate for use in other collisionally dominated plasmas, up to densities of $10^{15}$~cm$^{-3}$. Following a description in Sect.~\ref{sec:methods} of the methods used to modify the coronal approximation and the rates incorporated into the modelling, results from new electron impact ionisation (EII) calculations for all of the charge states of oxygen are presented in Sect.~\ref{sec:results}. The same section also illustrates how the ion populations alter with density as a result of the new modelling. Following this, in Sect.~\ref{sec:obs}, the new ion populations are tested by calculating theoretical intensities of many oxygen lines formed in the solar TR to assess how they compare with observations. A short conclusion is given at the end.

\section{Methods}
\label{sec:methods}

The population $N^z_{i}$ (for each temperature) of an ion with charge $+z$ in a level $i$ can be obtained from:

\begin{equation}
\eqalign{
	\frac{dN^{z}_{i}}{dt} = & \sum_{j}C^{z}_{ij}N^{z}_{j} + \sum_{j}S^{z-1}_{ij}N^{z-1}_{j} + \sum_{j}R^{z+1}_{ij}N^{z+1}_{j} \cr
	& - \sum_{j}C^{z}_{ji}N^{z}_{i} - \sum_{j}S^{z}_{ji}N^{z}_{i} - \sum_{j}R^{z}_{ji}N^{z}_{i} ~, \cr}
\end{equation}

\noindent where $C^{z}_{ij}$ represents the collisional-radiative matrix element for processes within an ion from level \textit{j} to level \textit{i}; $S^{z}_{ij}$ is the matrix element for ionisation processes from level \textit{j} of charge state $+z$ into level \textit{i} of the next higher charge state; and, $R^{z}_{ij}$ is the element for recombination processes out of level \textit{j} in charge state $+z$ into level \textit{i} of the next lower charge state. In addition, there is the normalisation condition that the total ion populations should be equal to the elemental abundance: $N(X)=\sum_z N^{z}$. 

The rates for collisional processes involving free electrons are dependent on both density and temperature, whereas radiative decay and photo-induced processes are independent of both of these quantities. Consequently, transitions connecting levels within an ion and between ions are related and level populations will alter as density and temperature change. In the solar corona where higher temperatures and lower densities prevail, charge states typically remain in the ground state, since radiative rates are fast and electron impact excitation (EIE) rates are slow. The plasma dynamics are slow enough that state steady equilibrium may also be assumed. The charge state distribution is, therefore, determined by ionisation and recombination rates between ground levels in an ionisation equilibrium, $dN^z/dt = 0$. Additionally, since photo-induced processes have negligible influence on the ion populations, and ionisation and recombination rates which are important in the solar corona are all linearly dependent on density, the density cancels out of the equations and the ionisation equilibrium is independent of density. This is known as the coronal approximation, and is the most commonly used form of modelling in collisional plasmas; it is used in the \textsc{Chianti} database \citep{dere2019}, for example.

For the solar TR several of these assumptions break down. Principally, cooler temperatures mean the presence of lower charge states, which have a more complex structure. Excited levels close to the ground do not have strong, dipole radiative decays. If densities are high enough EIE rates are sufficiently fast to maintain a population in these excited levels, which are known as metastable. The metastable levels, being closer to the continuum, typically have faster ionisation rates than the ground level. Zero density recombination rate coefficients from the metastable levels, however, are usually slower. This is because dielectronic capture from the metastable levels always occurs into levels which can auto-ionise back into the ground level of the initial ion.

Effective ionisation and recombination rates are the sum of the rates out of an ion weighted by the relative population of each level. At zero density, all the ions of one species will be in the ground level, and the effective rate would be equal to the ground level rate out of the ion. As density increases in the plasma and the species is collisionally excited into a metastable level, the effective rate would alter depending on the degree to which the metastable level is populated and by how much the rate out of the metastable level differs from that of the ground. It means that, in typical scenarios, as density increases the effective ionisation rate out of an ion will increase and the effective recombination rate will decrease. Thus, the charge state distribution will be different when metastable levels are included in the modelling compared to coronal approximation modelling, if densities are high enough for metastable levels to be populated. This was noted by \cite{burton1971} as a possible reason for the discrepancy in Li- and Na-like ion intensities, and the effect on line emission was explored by \cite{summers1983}.

Another density dependent effect was first demonstrated by \cite{burgess1969}, and arises when DR takes place into highly excited states close to the ionisation limit. Collisional ionisation rates become sufficiently strong to prevent radiative decay, thus reducing the rate at which recombination takes place. The combined effect of metastable levels and suppression of DR will alter how ions form in the solar TR, and, with it, the predictions of line emission when compared to the coronal approximation.

Owing to the preponderance of the coronal approximation, much effort has been put towards finding accurate ionisation and recombination rates from the ground level, (see, e.g., \citealt{arnaud1985, dere2007}). Apart from Be-like ions, where ions in metastable levels are often present in experiments, the prior lack of attention given towards ionisation from excited levels has been addressed only more recently, for some elements, by M.S.~Pindzola, D.C.~Griffin and co-workers. They use these rates for the same purpose of providing more accurate collisional radiative (CR) modelling, \citep[see, e.g.,][]{ludlow2008, ballance2009}. The first part of the current work is to calculate EII rate coefficients for the ground and metastable levels which can be included in the modelling. The rest of this section will highlight the methods used to calculate them and the other rates incorporated into the CR modelling.

\subsection{Electron impact ionisation}
\label{sec:ionmethods}

The \textsc{Flexible Atomic Code} (FAC) \citep{gu2008} has been used in this work to calculate direct ionisation (DI) cross sections for ground and metastable levels for all ion stages of oxygen. The distorted wave (DW) approximation used by FAC has shown discrepancy with experiment and more accurate theory (such as convergent close coupling and R-matrix) for neutral and low charge states. In these cases, the total cross sections from FAC were adjusted to experiment by using the scaling of \cite{rost1997}. 

Adjusting the cross sections was carried out by finding, for each level in an ion, the maximum in the cross section and its corresponding energy. The cross section and final scattered electron energy were normalised by the maximum and energy at maximum respectively. The maximum in the experimental cross section and its corresponding energy were used to determine the ratio of these values to the values from FAC, (after the subtraction of the excitation--auto-ionisation contribution from the total cross section). The maxima in the cross sections from FAC and their corresponding energies, for all levels, were adjusted by this ratio; the normalised cross sections and energies were then de-scaled to give new, adjusted cross sections. By this method, the shape of the original cross sections calculated by FAC are retained. This process assumes that the experimental results were for ions in their ground state, and that the cross sections calculated by FAC for the excited states were affected by the same extent as the ground state was. It is, therefore, noted that the results for the excited levels which have been adjusted have an additional degree of uncertainty.

For excitation--auto-ionisation (EA), \textsc{Autostructure} \citep[AS,][]{badnell2011} was used for all rates required in the calculation: EIE, auto-ionisation and radiative decay. For many of the ions, R-matrix collision rates are available. They use the same structure, but have the benefit of resonance contributions and close coupling for greater accuracy. Consequently, these rates are used where possible. The \textsc{ADASEXJ}\footnote{amdpp.phys.strath.ac.uk/autos/} post-processor was used to derive Maxwellian rate coefficients when the distorted wave collision strengths from \textsc{Autostructure} were required.

Both sets of rate coefficients are available online in electronic form at CDS, and are resolved by both initial and final level of the ions, except where scaling has been used, for which totals only are given. Total rate coefficients from each initial level are also included because, in some cases, there are contributions to the totals from transitions which are not included in those resolved by level. For ion stages where no metastable levels exist rate coefficients to and from one or two excited levels have been included in some cases.

\subsection{Recombination}
\label{sec:recmethods}

Collisional radiative modelling which includes density effects requires all the excited states to be included which make a noticeable contribution to the total recombination rate into an ion. For the TR ions of oxygen, at their formation temperatures, the DR rate coefficients into levels with $n>100$ still often contribute to the DR rate coefficients. Since not only would these rates be required for the model, but also the ionisation, excitation and radiative decay rates associated with transitions in and out of every level, the task is complex and, instead, has been approximated at this stage. To give an estimate of how the highly excited levels influence the ionisation equilibrium, the suppression of DR with density has been estimated in a similar fashion to other works, as, for example, in \cite{judge1995} and \cite{nikolic2013,nikolic2018}.

The estimate is carried out by using the ratio of the \cite{summers1974} effective recombination rates for the density being modelled to the rate of the lowest density ($10^4$ cm$^{-3}$), for each ion for which suppression is shown to affect recombination. The DR rate from both the ground and metastable levels is then multiplied by this ratio. The suppression factor is applied to the recombination rates from metastable levels because they show the same tendency for recombination to take place into Rydberg levels. Modelling the ions in this way means that partial DR rates into every level of the recombined ion are not required. The total rates, which are the sum over all the final states in the recombined ion, are used in this model for each of the ground and metastable levels. Owing to the nature of this modelling, the excited levels will only be populated by collisions, and no effects will arise from recombination enhancing their populations either directly or through cascades. This modelling is appropriate for studying just collisionally excited lines. The method estimates only the suppression with density of recombination into the Rydberg levels, and not the reduction in effective recombination rates caused when metastable levels become populated. This is because \cite{burgess1969} and \cite{summers1974} did not have metastable levels present in their modelling. Thus, there will not be a ``double counting'' of the suppression effects.

\cite{nikolic2013, nikolic2018} state that the method should be used only to indicate which ions are affected by suppression of DR. They say it is not a substitute for CR modelling in which Rydberg levels are included, which would reproduce the suppression of DR self-consistently. That is the aim in this paper: to determine how much DR suppression from the Rydberg levels affects the modelling of the oxygen ions, and whether full modelling is required. The approximation does assume that the current DR rates would be affected in the same way as the Summers recombination rates, and also that DR from the metastable levels would be suppressed by electron collisions in the same way as the ground level. The suppression factors were calculated using the Summers data made available on the OPEN-ADAS website\footnote{open.adas.ac.uk}; the effective recombination rates may be found in the ADF11 data class, in files with the prefix `\texttt{acd74}'.

Total radiative recombination (RR) rates may also be used, but in this case it is because nearly all recombination occurs into the ground complex for RR. For this reason, no suppression factors are required for RR rates. It was shown by \cite{dufresne2019} that level populations of carbon were altered by no more than 0.2\% using total RR rates, compared to using partial RR rates.

Much work has been done by N.R.~Badnell and co-workers to improve recombination rates for a wide range of ions. The total rates produced by them for each of the ground and metastable levels of oxygen have been incorporated into the model. Specifically, the RR rate coefficients used in the modelling all come from \cite{badnell2006}, while the DR rates come from the following sources: for \ion{O}{ii} \cite{mitnik2004}, \ion{O}{iii} \cite{zatsarinny2005}, \ion{O}{iv} \cite{altun2004}, \ion{O}{v} \cite{colgan2003}, \ion{O}{vi} \cite{colgan2004}, \ion{O}{vii} \cite{bautista2007}, and \ion{O}{viii} \cite{badnell2006dr}.

\subsection{Internal transition rates}
\label{sec:intmethods}

All of the radiative decay and proton and electron (de-)excitation rates were imported from \textsc{Chianti} v.9 \citep{dere2019}, with the exception of \ion{O}{i}, since \textsc{Chianti} includes only seven levels. For \ion{O}{i} a model in $LS$-coupling was built, using radiative decay rates from AS and EIE cross sections from \cite{tayal2016b}. The EIE rate coefficients were calculated from the cross sections using an integration routine made available by Paul Barklem\footnote{github.com/barklem/libpub}. For the remaining ions models were built in intermediate coupling. The original sources for the data in \textsc{Chianti} are: for \ion{O}{ii} \cite{tayal2007}, \ion{O}{iii} \cite{tachiev2001} and \cite{aggarwal1983}, \ion{O}{iv} \cite{liang2012} and \cite{correge2004}, \ion{O}{v} \citet[][which will appear in a future release of \textsc{Chianti}]{fernandez2014}, and \ion{O}{vi} \cite{whiteford2001}.

\section{Results}
\label{sec:results}

\subsection{Electron impact ionisation}
\label{sec:atmresults}

\cite{bell1983} reviewed the available theoretical and experimental cross sections for all ions of the elements from hydrogen to oxygen, to provide a set of recommended cross sections which could be used for modelling plasmas both in fusion and astrophysics. This provided a standard for the time, and many of the experiments selected are still used as reference works. \cite{dere2007} used FAC to calculate cross sections for the ground levels of all ions of hydrogen up to zinc. Thorough checks with experiment were made, and cross sections were occasionally adjusted to agree with experiment, especially for low charge ions. The results were incorporated into the \textsc{Chianti} database. Both these works provide convenient references because they include every ion of oxygen. Caution is used, however, for ions where cross sections are based on experiments which may include ions in metastable levels. Using such a result for the ground level cross section will produce an overestimate, changing both the zero density ion balance and how it alters with density. A review is given now of the cross sections calculated for this work and the comparison with experiment and theory.

\subsubsection{\ion{O}{i}}

The unsuitability of the DW method for neutral atoms is clearly seen for \ion{O}{i}, for which the cross section is 70\% higher than the experiment by \cite{brook1978}. Other methods available in FAC, the Coulomb-Born and binary encounter dipole approximations, are substantially higher than this. To provide rates for the CR model the total ionisation cross sections for the ground and metastable terms from \cite{tayal2016b}, calculated using the non-perturbative, B-spline R-Matrix (RM) method, have been used. The ground term cross section is in very good agreement with Brook et al. The cross sections are, however, only given up to an incident electron energy of 260eV, which is insufficiently low to calculate rate coefficients for the temperatures required in the CR model. To circumvent this, the cross sections are extrapolated using the high energy $\ln(u)/u$ scaling law for ionisation, where $u$ is the incident electron energy divided by the ionisation potential. Testing this method on other cross sections for which the high energy behaviour is known shows that the rate coefficients derived from extrapolated cross sections are identical up to $10^6$~K, and above that diverge by 10-20\% at the most. For \ion{O}{i} this is sufficient because it is completely ionised well below $10^5$~K. The rate coefficients are in $LS$-coupling, the same scheme in which the \ion{O}{i} model is built. Since they are total ionisation rates, they are posted to the ground level of the next charge state, where internal transitions quickly re-establish equilibrium level populations in the ion, in much the same way as discussed for recombination in Sect.~\ref{sec:recmethods}.

\subsubsection{\ion{O}{ii}}

\begin{figure}
	\centering
	\includegraphics[width=9.3cm]{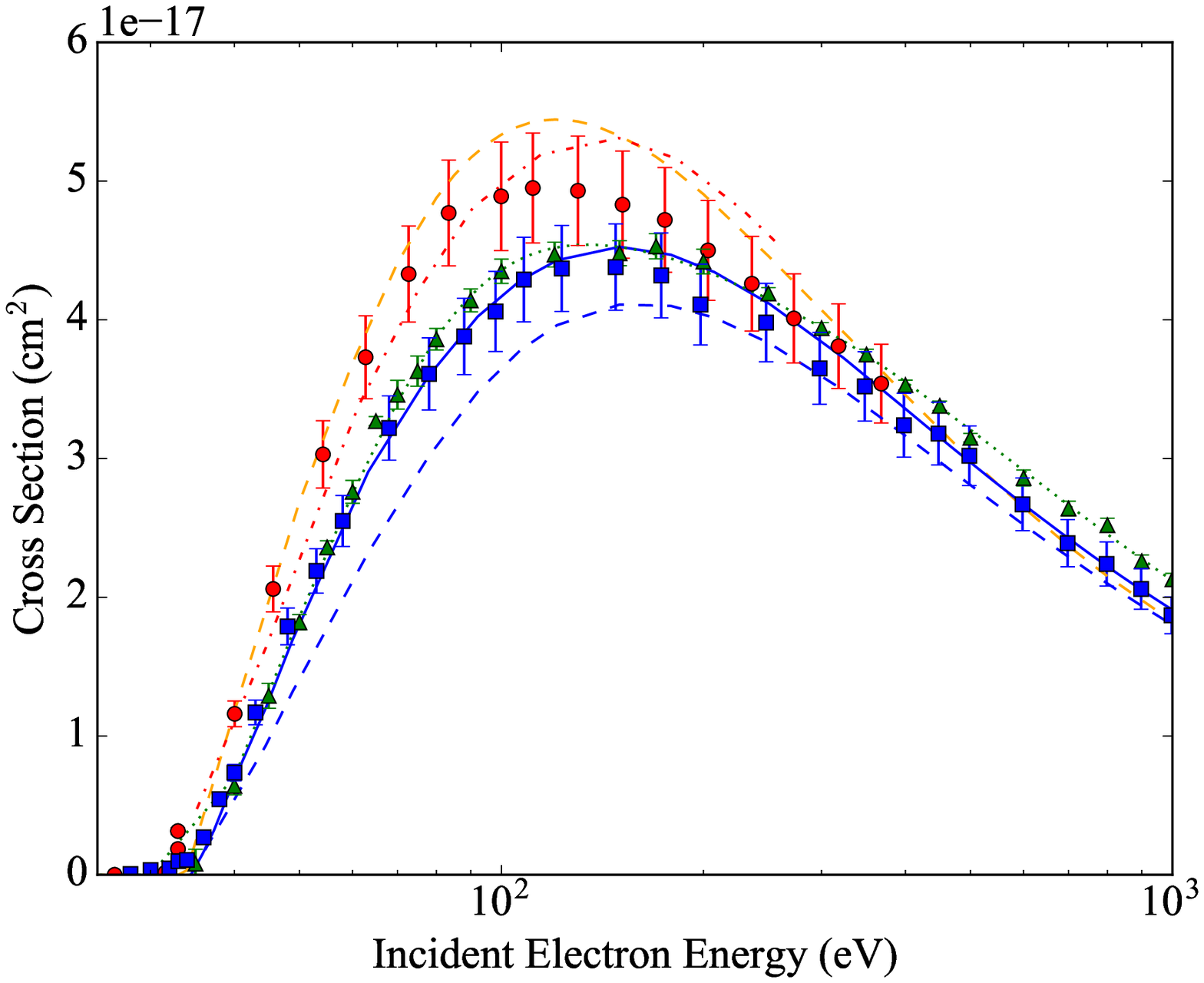}
	\caption[width=1.0\linewidth]{Total collisional ionisation cross section for \ion{O}{ii} ground level; blue solid line - this work (scaled), blue dashed - this work (scaled DI only), orange dashed - this work (unscaled DI only), green dotted - Dere, red dash-dotted - Loch et al., red circles - Loch et al. expt, blue squares - Aitken et al. expt, green triangles - Yamada et al. expt.}
	\label{fig:o2ionisgrd}
\end{figure}

\begin{figure}
	\centering
	\includegraphics[width=9.3cm]{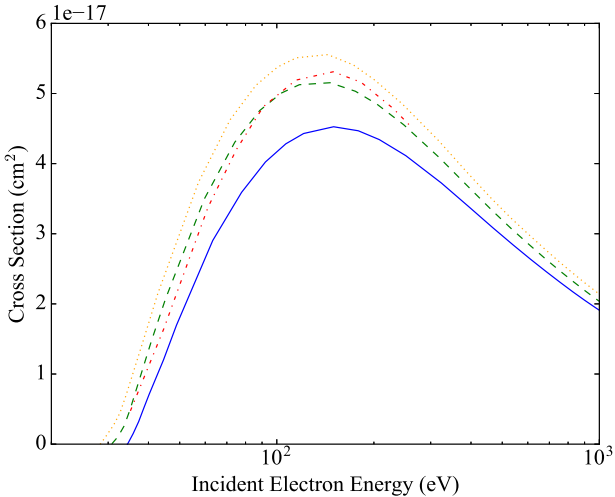}
	\caption[width=1.0\linewidth]{Total collisional ionisation cross sections for the \ion{O}{ii} $2s^22p^3$ ground configuration; blue solid line - this work $^4S$ ground term (scaled), green dashed - this work $^2D$ metastable term (scaled), orange dotted - this work $^2P$ metastable term (scaled), red dash-dotted - Loch et al. time dependent, close coupling, configuration average.}
	\label{fig:o2ionisterms}
\end{figure}

For EII of \ion{O}{ii}, the FAC cross sections are within the uncertainties of the \cite{hamdan1978} and \cite{loch2003} experiments, but 25\% higher than the \cite{aitken1971b} and \cite{yamada1988} experiments, which are in close agreement with each other. This, however, is before the addition of EA. As will be seen for the other ion stages, the Loch et al. experiment is often noticeably above other experiments, but for those ions obvious readings below threshold are present. With regards to theory, the FAC results peak a few per cent above the time-dependent close coupling results of Loch et al., although the FAC cross section peaks at lower electron impact energy and it does not include EA contributions, whereas the Loch et al. results do. If the rate coefficients are computed with the unadjusted cross sections, even before the addition of EA, they are 50\% higher than those recommended by \cite{bell1983} and 80\% higher than those of \cite{dere2007} at its peak formation temperature of 50,000~K.

Consequently, the cross sections from FAC for ground and metastable levels were scaled according to \cite{rost1997}, as described in Sect.~\ref{sec:ionmethods}, and adjusted to bring them into reasonable agreement with experiment. The \cite{aitken1971b} and \cite{yamada1988} experiments were used as the reference for the adjusting of the ground level cross section. The result for the ground state are shown in Fig.~\ref{fig:o2ionisgrd}. It can be seen how well the overall shape of the scaled cross section reproduces experiment, including at energies close to the threshold. Figure~\ref{fig:o2ionisterms} gives a comparison of the adjusted cross sections for all the terms in the ground configuration with the time dependent, close coupling results of \cite{loch2003}, which are calculated in the configuration average scheme. Since the highest and lowest terms from this work lie within their values, it highlights how the metastable terms have been effectively scaled using the \cite{rost1997} method, especially given that the only values adjusted are the cross section maxima and corresponding energies.

EA cross sections and rate coefficients were calculated using \textsc{Autostructure} DW excitation data, for which comparison of the thermally averaged collision strengths for individual transitions showed agreement of about 10\% with those calculated by \cite{kisielius2009}. Apart from a few of the higher $l$ states in the $2s^22p^2~3l,4l,5l$ configurations, all of the auto-ionising states contributing were from the $2s2p^3~4l,5l$ configurations. EA contributes 15\% to the total ionisation cross section of the ground term and 25\% to the metastable terms.

\subsubsection{\ion{O}{iii}}

\begin{figure}
	\centering
	\includegraphics[width=9.3cm]{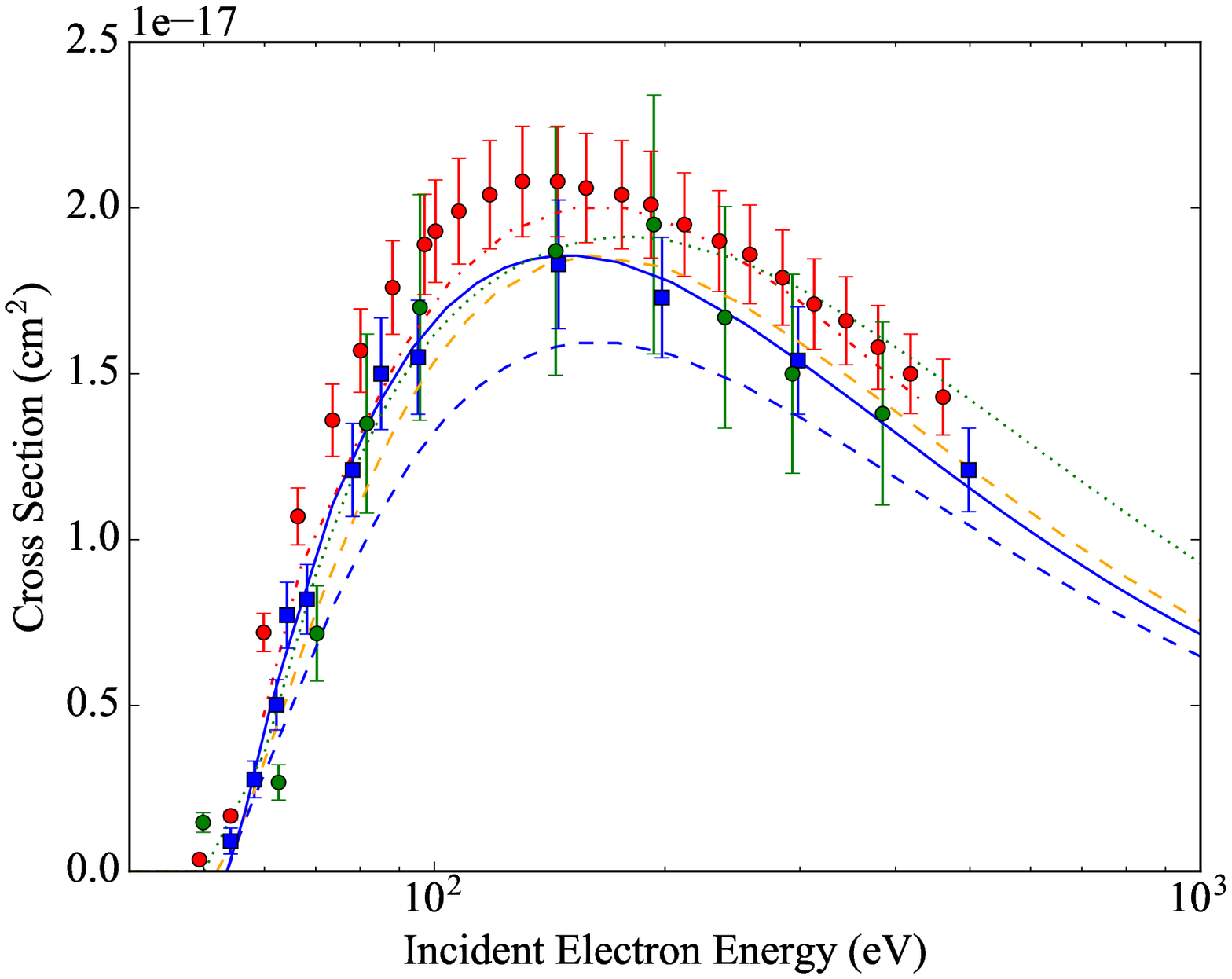}
	\caption[width=1.0\linewidth]{Total collisional ionisation cross section for \ion{O}{iii} ground state; blue solid line - this work (scaled), blue dashed - this work (scaled DI only), orange dashed - this work (unscaled DI only), green dotted - Dere, red dash-dotted - Loch et al., red circles - Loch et al. expt, green circles - Hamdan et al. expt, blue squares - Aitken et al. expt.}
	\label{fig:o3ionisgrd}
\end{figure}

Ionisation of \ion{O}{iii} shows closer agreement with other works than was the case for the lower charge states. The DI cross section is within a few per cent of the \cite{dere2007} DI cross section up to the peak, but drops more quickly at higher energies. It agrees well with the \cite{hamdan1978} and \cite{aitken1971b} experiments, but, again, this is before the addition of EA. Once this is added, the results are above all the other works. As a result, the cross sections were scaled and the cross section maxima adjusted by the ratio of the FAC cross section maximum for the ground level to the maximum of the \cite{aitken1971b} experiment. The energies were not adjusted. EA cross sections were calculated using the DW approximation in AS, and included excitations to $2s2p^2~4l,5l$ and $2p^3~3l,4l$ for all $l$ states above the ionisation threshold, beginning with $2s2p^2~4p~^{3}P$. The $2p^3~3l,4l$ levels are important for EA from the $2s2p^{3}~^{5}S$ level, which becomes populated at TR densities, because it requires excitation of just a single $2s$ electron. The final results are shown in Fig.~\ref{fig:o3ionisgrd}.

Regarding the rate coefficients, \cite{mao2020} has very recently released a new R-Matrix calculation containing many auto-ionising states. Carrying out a DW run using the same scaling parameters and configurations produced rate coefficients which are 50-60\% below the RM ones. Because of the variability in the DW rate coefficients for the lower charge states and they do not include resonance contributions, the RM results have been used. They were topped up with DW excitation to the $2s2p^2~5l$ configurations, which Mao et al. did not include, but which become important above $10^5$~K for EA.

\subsubsection{\ion{O}{iv}}

\begin{figure}
	\centering
	\includegraphics[width=9.3cm]{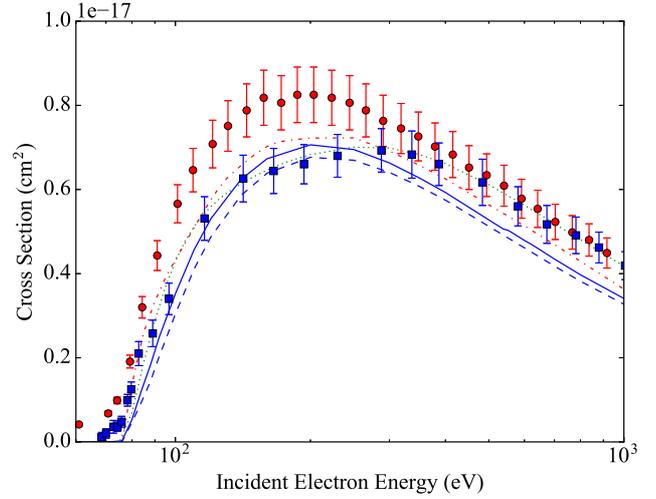}
	\caption[width=1.0\linewidth]{Total collisional ionisation cross section for \ion{O}{iv} with 16\% metastable population; blue solid line - this work, blue dashed - this work (DI only), green dotted - Dere (ground only), red dash-dotted - Loch et al., red circles - Loch et al. expt, blue squares - Crandall et al. expt.}
	\label{fig:o4ionisgrdmeta}
\end{figure}

Comparison of the theoretical cross sections of this work, \cite{loch2003} and \cite{dere2007} for the ground level of \ion{O}{iv} shows they agree to within 10\% of each other. The Dere cross section, however, does not include EA contributions. For the $2s2p^{2}~^{4}P$ metastable level, the total cross section is 15\% higher than the configuration average DW results of Loch et al., which would be a little lower, furthermore, if they were resolved for this term. Comparison with experiment shows that the total cross section is in good agreement with \cite{crandall1979} around the peak, although it is lower near threshold and at high energies. Crandall et al. claim a metastable population of approximately 16\% was present, which is consistent with expected populations in medium density plasmas. Blending the theoretical cross sections of this work to reflect this metastable population, and the same also for the cross sections of Loch et al., brings slightly improved agreement with experiment, (Fig.~\ref{fig:o4ionisgrdmeta}). For DI from the metastable term, FAC calculates a cross section to the ground state of \ion{O}{v} only for the $^{4}P_{1/2}$ level, and this is orders of magnitude lower than the cross section to the \ion{O}{v} metastable term. This means the DI threshold for the $^{4}P$ term is effectively 78.7eV, which is almost the same as the DI threshold for the ground term. The cross section in the \cite{crandall1979} experiment, however, has a noticeable cross section starting at 68.6eV, the metastable threshold to the ground in \ion{O}{v}, which FAC does not reproduce. The difference near threshold does not go on to affect the line emission predicted for this ion, principally because the metastable levels have a small population at TR densities.

The experimental results from Loch et al. lie above their theoretical results for the metastable configuration cross section. They mention possible contamination in the beam for \ion{O}{v} as a reason for below threshold values, and a linear offset in the results is speculated by \cite{fogle2008} as another explanation for the higher readings in the \ion{O}{v} Loch et al. experiment. Perhaps the same explanations extend to \ion{O}{iv}, as well.

With regards to EA, almost all of the $2p^{2}~3l,4l$ configurations lie above the threshold. These, however, only require excitation of a single electron from the metastable term and so, again, make a notable contribution to the EA rate coefficients. Furthermore, all of the $2s2p~4l$ configurations lie above the threshold of the metastable levels in \ion{O}{v}. Consequently, EA from the metastable levels in \ion{O}{iv} into the metastable levels of \ion{O}{v} contributes more to the rate coefficients above 100,000~K than EA into the ground term. The DW results still showed some variability depending on the structure, and the R-Matrix EA rate coefficients, using the data from \cite{liang2012}, showed an increase of over a third for the ground level in a like-for-like calculation, although there was only a 15\% increase for the metastable levels. As a result, the RM data were used, but DW contributions had to be added for the $2p^{2}~4l$ configurations. Inner shell excitation was also included, but these only contribute above $10^6$ K.

\subsubsection{\ion{O}{v}}

\begin{figure}
	\centering
	\includegraphics[width=9.3cm]{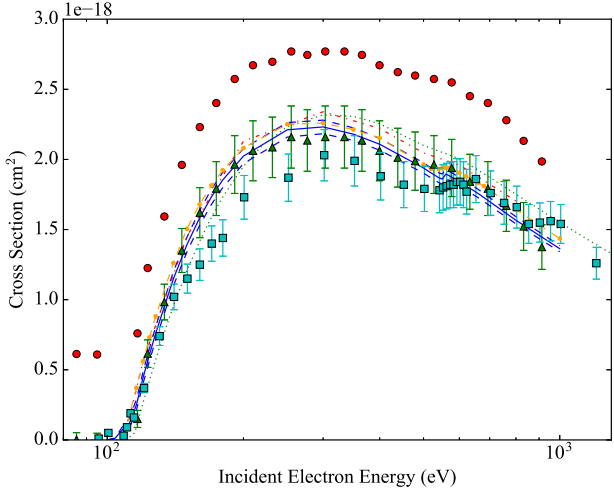}
	\caption[width=1.0\linewidth]{Total collisional ionisation cross section for \ion{O}{v} with 24\% metastable population; blue solid line - this work, blue dashed - difference in cross section owing to metastable population uncertainty, green dotted - Dere (ground only), red dash-dotted - Loch et al., orange dashed with circles - Fogle et al., red circles - Loch et al. expt, green triangles - Loch et al. expt (adjusted), light blue squares - Fogle et al. expt.}
	\label{fig:o5ionisgrdmeta}
\end{figure}

All the theoretical work surveyed are in close agreement for this ion, for which the theoretical cross sections from \cite{younger1981be}, \cite{loch2003} and \cite{fogle2008} are within 5\% of this work for the ground level. EA only contributes a small fraction to the cross section for this level because two electron excitation is required to reach the auto-ionising states. \cite{dere2007} cross sections match those from \cite{bell1983}, which are 15\% higher than this work. Owing to the high proportion of ions in metastable levels being present in experiments, more attention has been given in published work to ionisation from the metastable term than for other ions. Comparisons of the theoretical cross sections for the $2s2p~^{3}P$ term show good overall agreement. Fogle et al. is 5\% higher at the peak and Loch et al. is 10\% higher, while Younger is in very close agreement, although Younger does not include EA. Even though only single excitation of the $2s$ electron is required for EA from this term, the rate coefficients at $10^5$ K produced using the RM data of \cite{fernandez2014} were 50\% higher than those from DW calculations. To ensure the greatest accuracy, the former ones were used for the model. The total rate coefficients produced in this work are all within 5\% of those given by Dere, Bell et al. and Fogle et al..

Experiments have produced a wide variety of results and comparison is difficult because it is necessary to know what proportion of the ions are in metastable levels. \cite{crandall1979}, \cite{falk1983} and \cite{loch2003} are all close to each other. The cross section in Falk et al. is higher than the only theoretical result for the metastable level at the time. Consequently, they estimate the metastable population to be 90\%, which seems very high. Loch et al. note the significant below threshold behaviour in their experiment and speculate that an impurity may be present in the beam. Fogle et al., meanwhile, hypothesise that the Loch et al. values below threshold result from a linear offset in the measurements. Furthermore, they state the same type of equipment is used for their experiment as is used by Loch et al., and thus a similar proportion of metastable ions should be present in both experiments. They reduce the Loch et al. measurements by this offset and show that, for the most part, the two experiments lie within the experimental uncertainties of each other. If, alternatively, as Loch et al. suggest, their experiment contains an impurity which ionises at lower energy, it would be expected that at higher energies the impurity would contribute less to the cross section than it contributes below the \ion{O}{v} threshold. The offset would then be less at higher energies, and this may explain why the two experiments become closer in the high energy region.

Since Fogle et al. claim they know the population of ions in metastable states within a reasonable proportion, comparison with that experiment is made, (Fig.~\ref{fig:o5ionisgrdmeta}). The cross section for the ground and metastable levels weighted by their experimental ion populations (76\% ground, 24\% metastable) is in good agreement with the experiment, particularly near the threshold and at higher energies. Interestingly, the area where the cross section from this work lies outside the experimental uncertainty is at the energies just below the peak, where their experiment differs most with the adjusted Loch et al. results.

\subsubsection{\ion{O}{vi}}

The DI cross section from this work is 5\% lower than \cite{younger1981high} and 10\% lower than \cite{dere2007}. The cross section from Younger is obtained from the scaled cross section for the isoelectronic sequence, while Dere uses FAC. Since FAC is used for the present work, also, and tests with different optimisations and configurations produces the same results, perhaps a possible explanation is that a different version of FAC may have been used here (v. 1.1.4) than used by Dere. Inner shell EA makes a notable contribution to the cross section and has been clearly identified by experiment. The calculated EA cross section is 25\% higher than \cite{dere2007}, but this will not affect the rate coefficients substantially because the EA threshold is so much higher than that of DI.

\begin{figure}
	\centering
	\includegraphics[width=9.3cm]{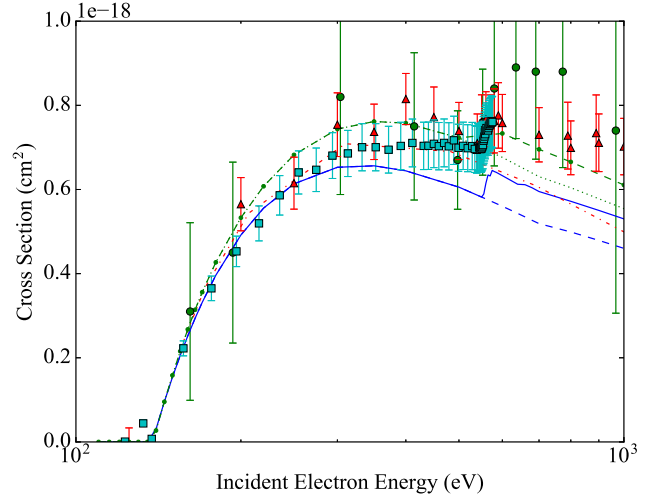}
	\caption[width=1.0\linewidth]{Total collisional ionisation cross section for \ion{O}{vi} ground level; blue solid line - this work, blue dashed - this work (DI only), green dashed with circles - Dere, green dotted - Dere (DI only), red dash-dotted - Younger, green circles - Crandall et al. expt, red triangles - Defrance et al. expt, light blue squares - Rinn et al. expt.}
	\label{fig:o6ionisgrd}
\end{figure}

In comparison with experiment, the computed cross section is favourable with the work of \cite{rinn1987}, lying close to it from threshold up to the peak, and thereafter the experiment is noticeably higher, as shown in Fig.~\ref{fig:o6ionisgrd}. The experiment of \cite{defrance1990} has few data points at energies below the peak in cross section, but this work does agree in that region. Above that, they lie higher than this work and their results are closer to Dere. \cite{crandall1979} has a large uncertainty, which encompass all of the theoretical results in most cases. The rate coefficients are in very good agreement with Dere and \cite{bell1983} for most of their range, and only differ at temperatures above $10^6$ K, varying by 10-20\%.

\subsubsection{\ion{O}{vii} and \ion{O}{viii}}

The various DW calculations for He-like \ion{O}{vii} are all in very good agreement with each other. There are only minor differences: the Bell et al. cross section is shifted to slightly higher energy, Dere is approximately 15\% higher and \cite{fontes1999} is greater at high energies. The \cite{younger1980he} results used for the comparison were obtained from the scaled cross sections given for the isoelectronic sequence. Bell et al. derived their recommended cross sections from data for \ion{B}{iv} in \cite{younger1980he} and scaling to \ion{O}{vii} using classical scaling laws. These differences translate into the rate coefficients from Bell et al. at lower temperatures being 60\% lower than this work and Dere, while at higher temperatures Dere is 10\% higher than this work and Bell et al. 

Bell et al. used a similar method to obtain their cross section for \ion{O}{viii}, which they derived by scaling the H-like \ion{C}{vi} given in \cite{younger1980h}. Their result at the peak in cross section is 15-20\% lower than the present work and Dere, both of which agree very well with the experiment of \cite{aichele1998}. Close to threshold there is good agreement for all of the cross sections. As a result, the rate coefficients of the present work agree very closely with Bell et al. and Dere.

\subsection{Collisional radiative model}
\label{sec:crmresults}

With all the ionisation rate coefficients just described, these are included level-by-level into the model, from the ground and metastable levels to as many levels in the next higher charge state as contribute to the total rate out of the ion. Ionisation from excited states higher than those which are metastable do not contribute towards shifting the ion balance, and have not been included. In addition, as noted in Sect. \ref{sec:recmethods}, the recombination rates are included from each of the ground and metastable levels as a single, total rate posted into the ground state of the recombined ion. They are not partial rates posted to every level in the recombined ion. Consequently, it means that all the other excited states which are included in the model exist just to establish the ground and metastable level populations, that is, no ionisation and recombination takes place from excited levels other than metastable. The rate equations for all levels of all the ions are solved at once for each temperature at a given density. To show the effect on the ion balance of the processes described in Sect.~\ref{sec:methods}, the model is run with each process included separately before being combined together to give the final ionisation equilibrium. Ion populations from the final ionisation equilibrium for a range of temperatures and densities will be made available at the CDS in the \textsc{Chianti} `\texttt{.ioneq}' file format.

\subsubsection{Effect of the metastable levels}
\label{sec:crmmeta}

This section considers how the presence of the metastable levels in the modelling alter effective rate coefficients out of an ion. This arises because the zero density rate coefficients of the metastable levels differ from those of the ground, as discussed in Sect.~\ref{sec:recmethods}. In this section, it is considered separately from the suppression of the DR process which occurs when recombination takes place into Rydberg levels and electron collisions prevent radiative stabilisation; that process is discussed in the next section.

To demonstrate the effect that the rate coefficients of the metastable levels have on the ionisation equilibrium, Tables \ref{tab:ionrate} and \ref{tab:recrate} show how the effective ionisation and recombination rate coefficients out of an ion alter once metastable levels become populated as the density in the plasma rises. \ion{O}{vi} shows how, if an ion has no metastable levels, the effective rate coefficients out of an ion do not change with density because the ion remains in the ground state. For all of the other transition region ions their rate coefficients change as higher densities raise the metastable level populations. The zero density ionisation rates differ from \textsc{Chianti} in this model because the new EII rate coefficients calculated for this work are used, rather than those of \cite{dere2007}; the effective recombination rate coefficients at zero density do not because the two models use the same data sources.

\begin{figure}
	\centering
	\includegraphics[width=8.7cm]{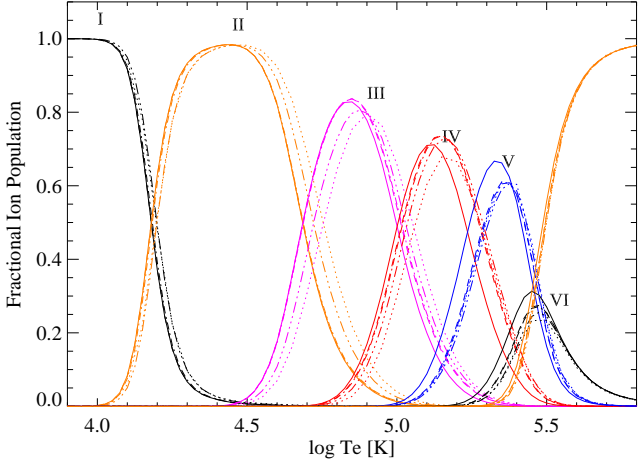}
	\caption{Effect with density of level resolved ionisation and recombination on the CR model for oxygen; dotted line - \textsc{Chianti}, dash-dot-dotted - this work at $10^4$ cm$^{-3}$ density, dash-dotted - $10^8$ cm$^{-3}$, dashed - $10^{10}$ cm$^{-3}$, solid - $10^{12}$ cm$^{-3}$. Individual charge states are highlighted by Roman numerals and different colours.}
	\label{fig:ocrmlr}
\end{figure}

The metastable levels in \ion{O}{i} all lie within the ground configuration, and so the energy differences are small between the terms and the ionisation rate coefficients similar. This produces little shift in the \ion{O}{i} populations as density increases, as illustrated in Fig.~\ref{fig:ocrmlr}. At the formation temperature of \ion{O}{i}, the first metastable term, $^1D$, reaches its highest population (less than 10\%) at an electron density of $10^{8}$ cm$^{-3}$, explaining the small shift in populations seen compared to the lowest density. The $^1S$ term reaches saturation at higher densities, but its population is less than 1\% and does not contribute noticeably to changing the \ion{O}{i} population. While it is true also for \ion{O}{ii} that the metastable terms are in the ground configuration and the ionisation rate coefficients similar, the populations of the metastable terms are substantially higher. At $45,000$K, the $^2D$ term reaches a peak population of 42\% at $10^{4}$ cm$^{-3}$ and the $^2P$ term reaches 14\% at $10^{8}$ cm$^{-3}$, which both contribute to a noticeable change in the populations of \ion{O}{ii} compared to the coronal approximation of \textsc{Chianti}.

\begin{table}
	\caption{Comparison of effective ionisation rate coefficients once metastable levels are included in the current work (in cm$^{-3}$~s$^{-1}$), at 100,000~K and various densities, plus a comparison with \textsc{Chianti}.} 
	\centering	
		\begin{tabular}{p{0.1in}rcccc}
			\hline\hline \noalign{\smallskip}
			Ion & Zero Density & $10^8$ cm$^{-3}$ & $10^{12}$ cm$^{-3}$ & \textsc{Chianti} \\
			\noalign{\smallskip}\hline\noalign{\smallskip}
			
				\ion{O}{i} & $4.6\times10^{-09}$ &  $5.0\times10^{-09}$ & $5.0\times10^{-09}$ & $4.8\times10^{-09}$ \\
				\ion{O}{ii} & $2.1\times10^{-10}$ &  $3.2\times10^{-10}$ & $3.2\times10^{-10}$ & $2.1\times10^{-10}$ \\
				\ion{O}{iii} & $1.8\times10^{-11}$ &  $2.1\times10^{-11}$ & $2.2\times10^{-11}$ & $1.4\times10^{-11}$ \\
				\ion{O}{iv} & $4.5\times10^{-13}$ &  $4.4\times10^{-13}$ & $5.0\times10^{-13}$ & $4.9\times10^{-13}$ \\
				\ion{O}{v} & $2.4\times10^{-15}$ &  $3.5\times10^{-15}$ & $7.6\times10^{-15}$ & $1.8\times10^{-15}$ \\
				\ion{O}{vi} & $4.2\times10^{-17}$ &  $4.2\times10^{-17}$ & $4.2\times10^{-17}$ & $4.2\times10^{-17}$ \\
			
			\noalign{\smallskip}\hline
		\end{tabular}
	\label{tab:ionrate}
\end{table}

\begin{table}
	\caption{Comparison of effective recombination rate coefficients once metastable levels are included in the current work (in cm$^{-3}$~s$^{-1}$), at 100,000~K and various densities, plus a comparison with \textsc{Chianti}. Suppression of DR from the Rydberg levels is not included. The effective rate coefficients are given for the initial, recombining ion.} 
	\centering	
		\begin{tabular}{p{0.1in}rcccc}
			\hline\hline \noalign{\smallskip}
			Ion & Zero Density & $10^8$ cm$^{-3}$ & $10^{12}$ cm$^{-3}$ & \textsc{Chianti} \\
			\noalign{\smallskip}\hline\noalign{\smallskip}
			
				\ion{O}{ii} & $4.1\times10^{-12}$ &  $2.2\times10^{-12}$ & $2.2\times10^{-12}$ & $4.1\times10^{-12}$ \\
				\ion{O}{iii} & $1.2\times10^{-11}$ &  $8.4\times10^{-12}$ & $8.0\times10^{-12}$ & $1.2\times10^{-11}$ \\
				\ion{O}{iv} & $3.0\times10^{-11}$ &  $2.7\times10^{-11}$ & $2.2\times10^{-11}$ & $3.0\times10^{-11}$ \\
				\ion{O}{v} & $5.0\times10^{-11}$ &  $4.5\times10^{-11}$ & $2.8\times10^{-11}$ & $5.0\times10^{-11}$ \\
				\ion{O}{vi} & $5.7\times10^{-11}$ &  $5.7\times10^{-11}$ & $5.7\times10^{-11}$ & $5.7\times10^{-11}$ \\
			
			\noalign{\smallskip}\hline
		\end{tabular}
	\label{tab:recrate}
\end{table}

The fractional population curves of \ion{O}{iii}, \ion{O}{iv} and \ion{O}{v} are closely related to each other because of the varying densities at which the metastable levels become populated. In \ion{O}{iii} the metastable terms in the ground configuration, $^1D$ and $^1S$, are fully populated by $10^{8}$ cm$^{-3}$. It is not until a density of $10^{12}$ cm$^{-3}$ that the metastable levels in \ion{O}{iv} and \ion{O}{v} become fully populated. Consequently, up to this density, \ion{O}{iv} and \ion{O}{v} show little change in their fractional populations, but the shift becomes particularly enhanced at the highest density shown. The effective ionisation rate coefficient of \ion{O}{iv} decreases slightly at $10^{8}$~cm$^{-3}$, compared to the zero density rate coefficient, because the $^2P_{3/2}$ level in the ground term becomes more populated and this has a rate coefficient 5\% smaller than the $^2P_{1/2}$ level. Clearly, the greatest change in the effective ionisation rate coefficients occur with \ion{O}{v}, explaining the increase in Li-like \ion{O}{vi}.

As discussed in Sect.~\ref{sec:methods}, the presence of metastable levels not only increases the effective ionisation rates out of the ions, but also reduces the effective recombination rates, when considering just the lower rate coefficients that the metastable levels have from the ground. This is also playing a part in shifting ion formation to lower temperatures seen here. For example, the DR rates of \ion{O}{v} into \ion{O}{iv} are at their peak around their formation temperatures. The metastable levels do not become fully populated until the density reaches $10^{12}$ cm$^{-3}$, which will reduce recombination into \ion{O}{iv}, the main contributor to the noticeable increase in population of \ion{O}{v} between 120,000-200,000K at this density.

\subsubsection{Suppression of dielectronic recombination}
\label{sec:crmdr}

\begin{figure}
	\centering
	\includegraphics[width=8.7cm]{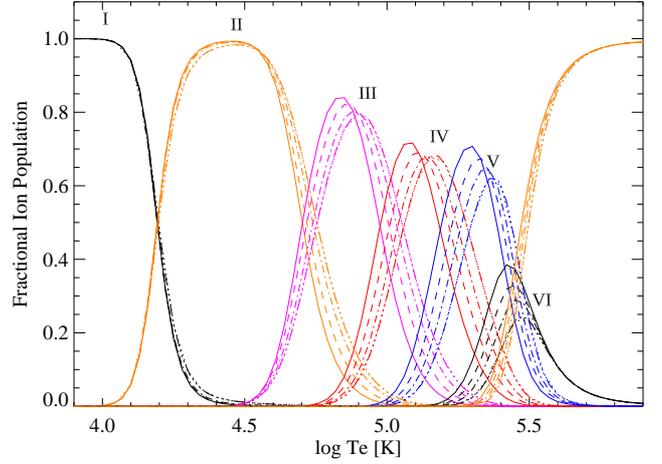}
	\caption{Effect with density of DR suppression on the coronal approximation model; dotted line - \textsc{Chianti}, dash-dot-dotted - this work at $10^4$ cm$^{-3}$ density, dash-dotted - $10^8$ cm$^{-3}$, dashed - $10^{10}$ cm$^{-3}$, solid - $10^{12}$ cm$^{-3}$. Individual charge states are highlighted by Roman numerals and different colours.}
	\label{fig:ocrmdr}
\end{figure}

This section considers how suppression of DR into the Rydberg levels at high densities affects the CR model. It is estimated by emulating how the effective recombination rates of \cite{summers1974} are suppressed with density. The impact on the charge state distribution can be best demonstrated by applying it to the coronal approximation model of \textsc{Chianti}, which uses total ionisation and recombination rates from the ground levels and will take no account of how the presence of metastable levels affect the distribution. The results are illustrated in Fig.~\ref{fig:ocrmdr}, and Table~\ref{tab:supprrate} shows how the effective rates in the CR model reduce as density increases.

\begin{table}
	\caption{Comparison of effective recombination rate coefficients in the coronal approximation when DR suppression from Rydberg levels is included (in cm$^{-3}$~s$^{-1}$), at 100,000~K and various densities, plus a comparison with \textsc{Chianti}. The effective rate coefficients are given for the initial, recombining ion.} 
	\centering	
		\begin{tabular}{p{0.1in}rcccc}
			\hline\hline \noalign{\smallskip}
			Ion & Zero Density & $10^8$ cm$^{-3}$ & $10^{12}$ cm$^{-3}$ & \textsc{Chianti} \\
			\noalign{\smallskip}\hline\noalign{\smallskip}
			
				\ion{O}{ii} & $4.1\times10^{-12}$ &  $1.8\times10^{-12}$ & $4.2\times10^{-13}$ & $4.1\times10^{-12}$ \\
				\ion{O}{iii} & $1.2\times10^{-11}$ &  $9.8\times10^{-12}$ & $3.4\times10^{-12}$ & $1.2\times10^{-11}$ \\
				\ion{O}{iv} & $3.0\times10^{-11}$ &  $2.3\times10^{-11}$ & $9.3\times10^{-12}$ & $3.0\times10^{-11}$ \\
				\ion{O}{v} & $5.0\times10^{-11}$ &  $3.3\times10^{-11}$ & $1.5\times10^{-11}$ & $5.0\times10^{-11}$ \\
				\ion{O}{vi} & $5.7\times10^{-11}$ &  $3.4\times10^{-11}$ & $1.5\times10^{-11}$ & $5.7\times10^{-11}$ \\
			
			\noalign{\smallskip}\hline
		\end{tabular}
	\label{tab:supprrate}
\end{table}

The DR rate coefficients into \ion{O}{i} are very low around its formation temperature; hence, there is little effect of suppression with density of recombination from \ion{O}{ii}. Conversely, the ground DR rate coefficients from each of \ion{O}{iii}--\ion{O}{vi} are at or near their peaks in their respective formation temperatures. Suppressing the rates according to the results of \cite{summers1974} indicates a close to linear decrease with density in the rates at these temperatures. The DR rates of \ion{O}{vii} are very small in TR conditions because, in order to initiate DR for this ion, it requires excitation of a $K$-shell electron. The thermal energy of electrons in the TR will be insufficient for this to occur. As a result, there will be little suppression of recombination from \ion{O}{vii}. Combined with the suppression of recombination from \ion{O}{vi} to \ion{O}{v}, there is a noticeable increase with density in the \ion{O}{vi} abundance.

\subsubsection{Final ionisation equilibrium}
\label{sec:crmioneq}

\begin{figure}
	\centering
	\includegraphics[width=8.7cm]{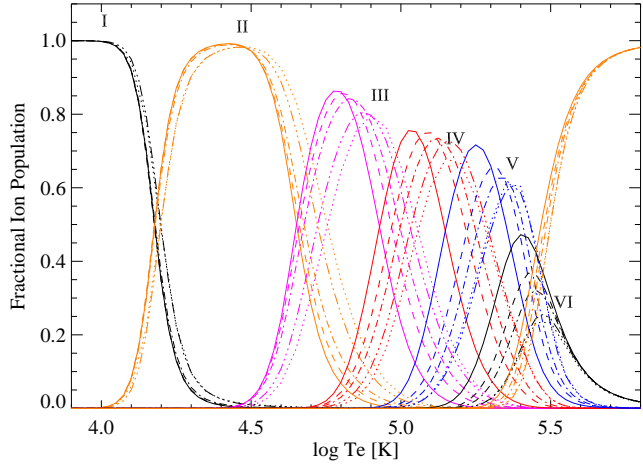}
	\caption{Final ionisation equilibrium of oxygen at various densities; dotted line - \textsc{Chianti}, dash-dot-dotted - this work at $10^4$ cm$^{-3}$ density, dash-dotted - $10^8$ cm$^{-3}$, dashed - $10^{10}$ cm$^{-3}$, solid - $10^{12}$ cm$^{-3}$. Individual charge states are highlighted by Roman numerals and different colours.}
	\label{fig:ocrmlrdr}
\end{figure}

Figure~\ref{fig:ocrmlrdr} shows the final, level resolved ionisation equilibrium for oxygen when combining density effects on the collisional ionisation and recombination processes discussed above. \ion{O}{ii}, while not increasing in abundance, shows a decrease in the temperature range of its formation, which could reduce the emission of lines forming at the higher end of its range. The peak abundances of \ion{O}{iii}, \ion{O}{iv} and \ion{O}{v} all increase by 10-15\% compared to the density independent modelling of \textsc{Chianti}. Similarly, in the same comparison, the peak formation temperatures of the TR ions all drop by 20-30\% in this model: \ion{O}{iii} from 79,000K to 66,000K, \ion{O}{iv} from 186,000K to 126,000K, and \ion{O}{v} down to 195,000K from 245,000K.

Comparisons of Figs~\ref{fig:ocrmlr}~and~\ref{fig:ocrmdr} at each density shows that it is suppression of DR which is producing the more significant shifts to larger populations and lower temperatures for the ions which are formed at higher temperatures in the solar TR. For the lower temperature ions it is ionisation from metastable levels which produces the greater change. Although in a CR model it is difficult to isolate all the factors which could explain this difference, some of it is attributable to how the atomic structure changes with atomic number along an isoelectronic sequence. As atomic number increases, radiative decay rates become stronger, requiring higher densities for the metastable levels to reach the same populations. This explains the metastable levels in \ion{O}{iii}--\ion{O}{v} not being highly populated until the density reaches $10^{12}$ cm$^{-3}$. Also, the energy separation between terms in the ground complex rise more slowly than the ionisation energy as atomic number increases along an isoelectronic sequence. Consequently, ionisation rate coefficients from metastable levels in more highly charged ions will become closer to those of the ground level, and will produce less shift in the charge state distribution. In contrast, the DR rate coefficients into \ion{O}{iii}--\ion{O}{vi} are at or near their peak in the solar TR, explaining the greater suppression of DR for these ions. Since, for oxygen, it requires densities higher than are present in the solar TR to noticeably alter the charge state distribution, the effect of density dependent modelling may become more important for emission from denser plasmas, such as those in solar and stellar active regions and flares.

\subsubsection{Comparison with other models}
\label{sec:crmcomp}

\begin{figure}
	\centering
	\includegraphics[width=8.7cm]{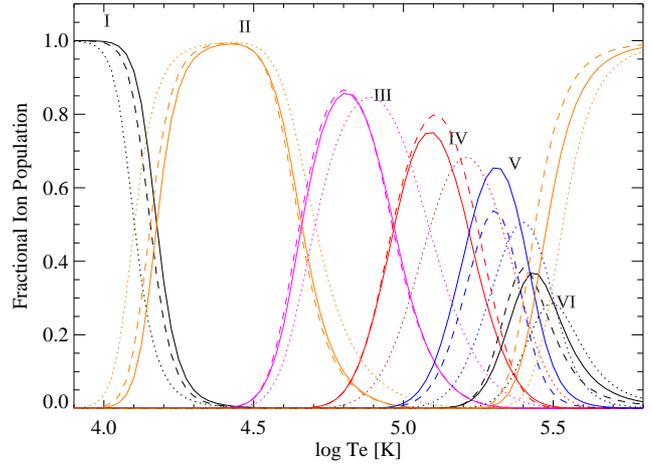}
	\caption{Comparison of the ionisation equilibrium of oxygen with other density dependent models at $10^{10}$ cm$^{-3}$ density; solid line - this work, dashed - OPEN-ADAS ADF11 1996 data class; dotted - Summers. Individual charge states are highlighted by Roman numerals and different colours.}
	\label{fig:ocrmcomp}
\end{figure}

The only results with which to compare this work with other density dependent CR models are those of \cite{summers1974} and from the Atomic Data and Analysis Structure (ADAS) consortium. The latter modelling is based on the original work by \cite{mcwhirter1984} and detailed in \cite{summers2006}. Effective ionisation and recombination rates are available, which make it possible to reconstruct the ionisation equilibrium. Effective rates from different types of modelling are available, but the ionisation equilibrium used as a comparison is the data from the ADF11 96 data class, because metastable levels were included in the modelling. The populations provided by the consortium (M. O'Mullane, private communication) are identical to the ion populations reconstructed from the OPEN-ADAS effective rates. Figure~\ref{fig:ocrmcomp} shows a comparison of the ionisation equilibria.

\cite{summers1974} used a hydrogenic model, which does not treat the effect of metastable levels, explaining why the results show less effect with density and are closer to the distribution shown in Fig.~\ref{fig:ocrmlrdr} for \textsc{Chianti}. Summers also used the approximations which were available at the time for rate coefficients, which could explain many of the differences and perhaps most especially for the difference with \ion{O}{i}, because this cannot be explained by either ionisation from metastable levels or DR suppression. 

Comparison of the current work and that of OPEN-ADAS shows how all of the charge states are noticeably affected by density in the TR. The populations of \ion{O}{i}-\ion{O}{iii} are the same for both works; the main differences lie in the more highly charged ions. Differences will arise, in part, from the ionisation rates. Inspection of the OPEN-ADAS ionisation rates shows that the rates from \cite{bell1983} are used for the ground. The results in Sect.~\ref{sec:atmresults} show that the ionisation rates used in this work are higher for \ion{O}{iv} than Bell et al., whereas they are lower for \ion{O}{v}, which explains part of the difference in the populations of those ions.

To check the differences between the ionisation equilibria of OPEN-ADAS and this work, the effective rates derived in both works were compared. For \ion{O}{i} the effective ionisation rates are very similar in both works, while the effective recombination rates into the ion from \ion{O}{ii} are 60\% lower in OPEN-ADAS, at the temperature where the ion has peak abundance. Similarly, the effective ionisation rates from \ion{O}{vi} are the same in both works and cannot account for the difference. The effective recombination rates into \ion{O}{vi} from \ion{O}{vii} are the same in both works at the peak in DR, which occurs at 3$\times$10$^6$~K, but not at lower temperatures, where the ion forms. Since there is also little DR suppression into both \ion{O}{i} and \ion{O}{vi}, it is clear that the primary difference between both works arises from the radiative recombination data. With regards to the \ion{O}{v} populations, the main cause of the difference arises in the effective recombination rates from \ion{O}{vi}. They are a factor of two lower in OPEN-ADAS, even at the lowest density, $10^8$ cm$^{-3}$. While the different radiative decay and EIE rates used in the two works would alter the metastable level populations, and thus change the effective ionisation and recombination rates, the differences in metastable populations are likely to be small and would not produce significant shifts in the ion populations.

\section{Comparison with observations}
\label{sec:obs}

As a way of testing how the new ionisation equilibrium fares against observations, it will be used to calculate line emission in the solar TR. The same calculations will also be carried out using the coronal approximation ion fractions from \textsc{Chianti}, to test how much the intensities are altered by density dependent effects. Tests on quiet Sun lines emitted by \ion{O}{iii}-\ion{O}{v} were carried out by \cite{doschek1999}, and they found discrepancies between predicted and observed intensities of more than a factor of two for \ion{O}{iv} and \ion{O}{v}. The same lines will be tested here, in addition to other oxygen lines, including those emitted by \ion{O}{ii} and \ion{O}{vi}.

\subsection{Method to calculate predicted intensities}

Doschek et al. did not give absolute intensities observed by the Solar Ultraviolet Measurements of Emitted Radiation (SUMER) instrument on board the Solar and Heliospheric Observatory (SOHO), instead expressing the results as ratios with the inter-combination lines emitted above the Lyman limit. Radiances for the same lines are given by \cite{warren2005}, which were also observed by SUMER during the same time period. When the data from Warren is expressed in the same ratios as Doschek et al., it shows that the observations given by Warren are within 10\% of those from Doschek et al. for \ion{O}{iii} and \ion{O}{iv} and within 20\% for \ion{O}{v}. These differences are well within the solar variability, especially when considering that lines more than 40\AA~apart are not observed simultaneously by SUMER. The observation of \cite{brekke1993} for the \ion{O}{iii} 1660.79\AA~line is also included, to be able to test an inter-combination line for this ion. The intensity from the quiet Sun region A data was selected because intensities from this region for the \ion{O}{iv} inter-combination lines tested here were within 15\% of Warren.

The main difference in methods to calculate predicted intensities is that Doschek et al. derived intensities by assuming isothermal conditions, that is, by assuming each line is emitted at the temperature of peak ion abundance. Here, differential emission measure (DEM) modelling, where the radiation is assumed to be emitted from a multithermal atmosphere, will be used to compute the intensities. To fit the DEM, many of the relatively strong lines with few blends used by \cite{dufresne2019} from \cite{vernazza1978} and \cite{wilhelm1998a} were chosen, except the lower resolution Skylab lines from \cite{vernazza1978} were replaced by measurements from Warren, if available. Some higher temperature lines from the Coronal Diagnostic Spectrometer (CDS), also on board SOHO, as reported by \cite{warren2005}, were included to help determine the DEM higher in the TR. No lines from Li-like and Na-like ions were used to fit the DEM. The same EIE and radiative decay data incorporated into the CR model were used to determine level populations. Ion populations were taken from this work for oxygen, \cite{dufresne2019} for carbon, and the default ones in \textsc{Chianti} for all other elements. To calculated intensities predicted by coronal approximation modelling the default ion balances in \textsc{Chianti} v.9 for all elements were employed. The photospheric abundances of \cite{asplund2009} were included. Checks against other photospheric abundances produced differences for most ions of only one or two per cent, but the Asplund et al. abundances improved the agreement for sulphur and silicon lines. A constant pressure of 3$\times$10$^{14}$ cm$^{-3}$ K$^{-1}$ was assumed, consistent with the model atmosphere given by \citet{avrett2008}. The DEM routine from \textsc{Chianti} v.9 was utilised.

\begin{table}
	\caption{Comparison of predicted and observed quiet Sun, oxygen radiances.} 
	\centering	
		\begin{tabular}{lrrcccc}
			\hline\hline \noalign{\smallskip}
			Ion & $\lambda_{\rm obs}$ & $I_{\rm obs}$ & $T^{\rm (1)}_{\rm eff}$ & $T^{\rm (2)}_{\rm eff}$ & $R^{\rm (1)}$ & $R^{\rm (2)}$ \\
			\noalign{\smallskip}\hline\noalign{\smallskip}
			
\ion{O}{ii} &  834.47 & 31.0 &  4.63 &  4.53 & 2.85 & 2.35 \\
\ion{O}{ii} &  832.76 & 12.9 &  4.63 &  4.53 & 2.28 & 1.88 \\
\ion{O}{ii} &  833.33 & 21.9 &  4.63 &  4.54 & 2.67 & 2.20 \\
\ion{O}{ii} &  718.50 & 14.6 &  4.69 &  4.62 & 2.86 & 1.66 \\
\noalign{\smallskip}
\ion{O}{iii} & 1660.79 & $19.3^a$ &  4.82 & 4.76 & 0.44 & 0.58 \\
\ion{O}{iii} &  835.29 & 78.6 &  4.88 &  4.82 & 1.01 & 1.13 \\
\ion{O}{iii} &  833.74 & 51.0 &  4.88 &  4.82 & 1.08 & 1.21 \\
\ion{O}{iii} &  835.10 & 11.7 &  4.89 &  4.82 & 1.20 & 1.34 \\
\ion{O}{iii} &  702.85 & 27.5 &  4.91 &  4.84 & 1.12 & 1.15 \\
\ion{O}{iii} &  703.85 & 43.5 &  4.92 &  4.85 & 1.12 & 1.16 \\
\ion{O}{iii} &  702.33 & 9.3 &  4.92 &  4.85 & 1.06 & 1.10 \\
\ion{O}{iii} &  599.56 & 35.7 &  4.97 &  4.89 & 0.97 & 0.89 \\
\ion{O}{iii} &  525.83 & 17.8 &  5.00 &  4.92 & 0.79 & 0.69 \\
\noalign{\smallskip}
\ion{O}{iv} & 1399.77 & 6.1 &  5.20 &  5.12 & 0.49 & 0.59 \\
\ion{O}{iv} & 1401.16 & 36.1 &  5.21 &  5.14 & 0.46 & 0.54 \\
\ion{O}{iv} &  787.71 & 58.5 &  5.24 &  5.17 & 1.04 & 1.08 \\
\ion{O}{iv} &  790.19 & 108.2 &  5.24 &  5.17 & 1.14 & 1.17 \\
\ion{O}{iv} &  554.10 & 40.1 &  5.25 &  5.19 & 1.09 & 1.03 \\
\ion{O}{iv} &  553.37 & 22.4 &  5.25 &  5.19 & 1.00 & 0.94 \\
\ion{O}{iv} &  555.29 & 24.5 &  5.25 &  5.19 & 0.92 & 0.86 \\
\ion{O}{iv} &  554.55 & 111.7 &  5.25 &  5.19 & 1.00 & 0.94 \\
\ion{O}{iv} &  608.38 & 17.7 &  5.25 &  5.21 & 1.06 & 1.03 \\
\noalign{\smallskip}
\ion{O}{v} & 1218.35 & 89.7 &  5.36 &  5.32 & 0.43 & 0.50 \\
\ion{O}{v} &  760.21 & 4.2 &  5.38 &  5.34 & 0.95 & 1.00 \\
\ion{O}{v} &  758.68 & 6.1 &  5.38 &  5.34 & 1.12 & 1.18 \\
\ion{O}{v} &  760.43 & 18.8 &  5.38 &  5.34 & 1.07 & 1.14 \\
\ion{O}{v} &  759.44 & 4.7 &  5.38 &  5.34 & 1.11 & 1.18 \\
\ion{O}{v} &  762.00 & 6.2 &  5.38 &  5.34 & 1.04 & 1.10 \\
\ion{O}{v} &  629.78 & 338.5 &  5.38 &  5.34 & 1.16 & 1.22 \\
\noalign{\smallskip}
\ion{O}{vi} & 1031.93 & 354.0 &  5.79 &  5.75 & 0.42 & 0.56 \\
\ion{O}{vi} & 1037.64 & 192.0 &  5.79 &  5.75 & 0.39 & 0.51 \\
        
			\noalign{\smallskip}\hline
			\noalign{\smallskip}
	\multicolumn{7}{p{0.9\columnwidth}}{\textbf{Notes.} Ion - principal ion emitting at observed wavelength $\lambda_{\rm obs}$ (\AA); $I_{\rm obs}$ - the measured radiance (ergs cm$^{-2}$ s$^{-1}$ sr$^{-1}$) from Warren, except for: a) from Brekke; $T_{\rm eff}$ - the effective temperature for each line (log values, in K), and $R$ - the ratio between the predicted and observed intensities using: (1) \textsc{Chianti} coronal approximation ion fractions, and (2) density dependent ion fractions for oxygen and carbon and \textsc{Chianti} for all other elements.}
		\end{tabular}
	\normalsize
	\label{tab:lines1}
\end{table}

\subsection{Results of DEM modelling}

The results of the DEM modelling for the oxygen lines are shown in Table~\ref{tab:lines1}, in which the final column, $R^{\rm (2)}$, gives the ratio of the theoretical intensities to the observed using the density dependent ion fractions, while the column $R^{\rm (1)}$ gives the ratio when coronal approximation ion fractions are used. The effective temperature, $T_{\rm eff}$, of a line is an average temperature more indicative of where a line is formed, which is different from both the temperature where the contribution function of a line has its maximum and where an ion has peak abundance. More details about the temperatures of lines and their differences can be found in, for example, \cite{delzanna2018}. Table~\ref{tab:linesdem} shows the same ratios for the lines used to fit the DEM. The lines for \ion{C}{i} from \cite{sandlin1986} and \ion{Fe}{xvi} from \cite{vernazza1978} in Table~\ref{tab:linesdem} are included in the modelling purely to keep the shape of the emission measure at the very lowest and highest temperatures.

\begin{figure}
	\centering
	\includegraphics[width=8.7cm]{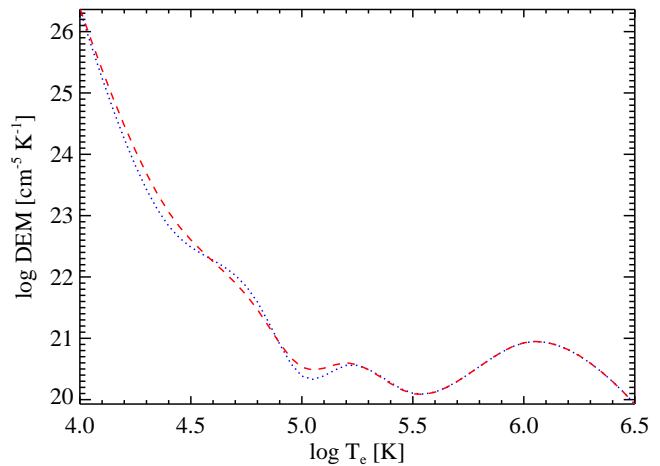}
	\caption{Comparison of the DEMs derived from modelling the lines in Table~\ref{tab:linesdem}; blue dotted line - using \textsc{Chianti} coronal approximation ion populations, red dashed - using ion populations from level resolved modelling for carbon and oxygen, and \textsc{Chianti} for all other elements.}
	\label{fig:dem}
\end{figure}

Figure~\ref{fig:dem} shows the DEM derived from the \textsc{Chianti} v.9 ion fractions and the DEM derived when the density dependent ion populations for carbon and oxygen are used. It would appear that further changes to the DEM could be expected if density dependent ion fractions for other elements are used to fit the DEM. The greater change in the DEM is at the lower temperature end of the solar TR, where the carbon ions form, rather than where most of the oxygen ions form. This explains why, of all the oxygen ions, the lines emitted by \ion{O}{ii} undergo the greatest change in predicted intensities compared to zero density modelling.

The DEM derived from the density dependent ion populations is lower in the region $log~T(K) = 4.6-4.9$, where the \ion{O}{ii} lines form, thus reducing their intensities. Furthermore, since the ion populations produced in density dependent modelling move to lower temperature, the contribution functions of lines which form at the higher end of the ion formation temperature will be reduced. This is another factor to explain why the higher temperature 718.50\AA~line decreases by 40\% compared to lines in the multiplet around 833\AA, which decrease by 20\%. With either type of CR modelling, there are significant discrepancies for the \ion{O}{ii} lines. Clearly, there are factors affecting its emission which ionisation equilibria with only electron collisional processes are unable to explain. This is not the case for many of the other lines which form around the same temperature, as shown in Table~\ref{tab:linesdem}. With density dependent ion populations for \ion{C}{ii} and \ion{C}{iii}, for instance, the ratios within each ion have been brought into better agreement with each other, and are closer to unity.

The DEM is not so steep around the region where the \ion{O}{iii}-\ion{O}{v} lines form. Consequently, the shift to lower formation temperatures demonstrated for these ions in this work does not produce as much change in the intensities compared to the \ion{O}{ii} lines. It cannot be determined at this stage whether the DEM is less affected in this region because the coronal approximation is sufficient to describe higher temperature ions, as discussed in Sect.~\ref{sec:methods}, or whether it is because there are fewer lines in that region modelled by the density dependent ion fractions.

For \ion{O}{iii} the predictions for the majority of the lines are in good agreement with observations. The main effect which may be seen in comparison to zero density modelling is that, again, the lower temperature lines, such as the 833\AA~multiplet, are marginally enhanced compared to zero density modelling, whereas the higher temperature lines are reduced, as seen for the 599.56\AA~and 525.83\AA~lines. The \ion{O}{iii} lines Doschek et al. modelled all show very good consistency here, which they also found. The 1660.79\AA~inter-combination line is furthest from observations, although the density dependent modelling has improved its predicted intensity by 32\%. Since the 525.83\AA~line comes from a more highly excited level than the others, there is a possibility the upper level population of this line could be enhanced in two ways which are not included in this model. There are just 46 levels for \ion{O}{iii} in the \textsc{Chianti} v.9 database, and, if more levels were included, the level could be enhanced by radiative cascades from higher levels following excitation. Alternatively, the population of this level could be increased through recombination directly into the level or from cascades following recombination into higher levels, both of which this model cannot simulate.

Both sets of CR modelling reproduce the solar emission very well for most of the \ion{O}{iv} lines, except for the inter-combination lines. Density dependent modelling improves the predictions for these lower temperature, inter-combination lines by 17-20\%, but observations show their emission is significantly stronger than the models predict. Doschek et al. used the 1401.16\AA~inter-combination line to normalise their intensities. If the same method was used here, it would cause an apparent inconsistency of almost a factor of two between observed and predicted intensities for all the other lines. This appears to be the main reason for the discrepancy in the Doschek et al. ratios. 

It is clear that the same argument holds for the \ion{O}{v} lines. They are all within reasonable agreement with observations except for the inter-combination 1218.35\AA~line. This, again, is predicted to be a factor of two below observations. The results from the density dependent modelling for this line does, however, show a 16\% improvement on zero density modelling, similar to the increase seen for the \ion{O}{iv} inter-combination lines.

Although the predicted intensities for the Li-like \ion{O}{vi} lines are both about a factor of two below observations, the results from the present work are an improvement of just over 30\% when compared to the ratios obtained from coronal approximation modelling. This is consistent with the improvement which \cite{doyle2005} predict would occur for this ion when density effects are added to the CR modelling.

With regards to the causes of the significant discrepancies between observed and theoretical intensities, \cite{doschek1999} propose that some of the differences they found could be due to inaccurate atomic data. They included resonant excitation because much of the EIE data at the time did not take it into account. The present work does include resonant excitation for all of the oxygen lines, and so that cannot account for the differences now. In \cite{doschek2004} an assessment of the widths of many inter-combination lines formed in the TR shows that the lines have narrower widths than other TR lines, suggesting the inter-combination lines form in different regions than allowed TR lines. By way of a further explanation, they suggest the lines may be enhanced by recombination into the upper levels of the inter-combination lines. This would account for higher intensities being observed than are predicted by the current work.

Another potential cause of the discrepancy could be in the observational data. \cite{doschek1999} normalised their lines to the inter-combination lines to remove uncertainty in case those below the Lyman limit were affected by absorption. They confirmed in their work that no evidence for this was apparent, and it is seen from the results here that lines used for fitting the DEM both above and below the Lyman limit are, generally, in good agreement with observations. Solar variability is another important factor to consider. Variations in the observations of \cite{doschek1999} between different dates can clearly be seen. Given the fact that the ratios are all normalised, it is hard to determine how much the absolute variations were. Variations in the normalising line could exaggerate, or indeed cancel out, changes in the other lines. For \ion{O}{iii}, which shows the least discrepancy in their modelling, the Doschek et al. normalised intensities vary over different dates by up to 50\%, while for \ion{O}{iv} the variation is up to 75\%. The intensities of the quiet Sun, \ion{O}{iv} inter-combination lines in \cite{sandlin1986} are about a factor of two stronger than those used here. The SUMER results of \cite{wilhelm1998a} and \cite{warren2005} often show differences of 40\% or more for the same lines. With this level of uncertainty, for a line which has a predicted to observed ratio of 0.50, the ratio may actually lie between 0.30 and 0.83. Using the Wilhelm et al. intensity of the \ion{O}{vi} 1037.64\AA~line in the modelling, for instance, results in a predicted to observed intensity ratio of 0.82.

\section{Conclusions}
\label{sec:concl}

As seen from the results of collisional direct ionisation calculations using the \textsc{Flexible Atomic Code} in this work and \cite{dufresne2019}, FAC produces cross sections which accurately reflect experiment for ions with a charge of $+3$ and higher. It is also noted that the FAC cross sections tend to peak at lower energy and drop more rapidly at high energies than experiment and other theory. When the results do not reflect experiment, the theoretical cross sections may be adjusted according to the scaling of \cite{rost1997}, by simply shifting the cross section maximum and its corresponding energy, and using the new values with the scaled cross sections. To the direct ionisation data must be added indirect ionisation, through the excitation--auto-ionisation process, particularly where EA occurs for outer shell electrons. It makes a significant contribution to the rate coefficients, which, in turn, alter the ionisation equilibrium.

In collisional radiative modelling, for the lighter elements modelled so far, the influence of the metastable levels makes almost as noticeable a contribution to the ion balance as suppression of dielectronic recombination. In order to provide a more accurate reflection of conditions in higher density plasmas, it should not be neglected. The focus of this work has been on processes which dominate in electron collisional plasmas. Other processes which may affect ion populations in the lower solar transition region, such as photo-excitation, photo-ionisation and charge transfer, have not been included in this model. They will be explored in future work, along with modelling which self-consistently determines suppression of dielectronic recombination for ions which have been shown to be particularly affected.

By comparing the results of the CR modelling with observations, it is seen that the ionisation equilibrium derived here accurately predicts the emission for the majority of oxygen lines observed in the solar transition region. The density dependent modelling has shown improvement in predicted line intensities over the zero density approximation by 15-40\% for \ion{O}{ii} lines, the inter-combination lines of \ion{O}{iii}-\ion{O}{v}, and Li-like \ion{O}{vi} lines. There, however, remain discrepancies of approximately a factor of two for all of these lines compared to observations. A certain amount of this difference could be explained by the variations in the intensities recorded in different works. Although density effects in plasmas dominated by electron collisions are not sufficient to fully describe conditions in more complex regions like the solar transition region, it is important that, to effectively make use of the data from Solar Orbiter and other missions, more complex modelling than the coronal approximation should be employed when interpreting emission from higher density plasmas.

\section*{acknowledgements}
	
	The authors would like to acknowledge: H.E. Mason for the helpful questions asked and guidance given during the course of the work; P.J. Storey for general discussions and for the insights given on the influence of atomic structure on transition rates and collisional radiative modelling; and, the reviewer for giving helpful suggestions, which improved the paper.

	Support by STFC (UK) via the consolidated grant of the DAMTP astrophysics group at the University of Cambridge and a Doctoral Training Programme Studentship are acknowledged, plus the support of a University of Cambridge Isaac Newton Studentship. NRB is funded by STFC Grant ST/R000743/1 with the University of Strathclyde. \
	
	Most of the atomic rates used in the present study were produced by the UK APAP network, funded by STFC via several grants to the University of Strathclyde. Acknowledgment is made of the use of the OPEN-ADAS database, maintained by the University of Strathclyde. \
	
	\textsc{Chianti} is a collaborative project involving George Mason University, the University of Michigan, the NASA Goddard Space Flight Centre (USA) and the University of Cambridge (UK). \

\bibliographystyle{mnras}

\bibliography{oioneq}

\appendix

\section{Lines used in the DEM modelling}

Table~\ref{tab:linesdem} lists the lines which were used to fit the DEM and the results for the ratios of predicted to observed intensities.

\begin{table}
	\caption{Comparison of predicted and observed quiet Sun radiances for lines used to fit the DEM.} 
	\centering	
		\begin{tabular}{lrrcccc}
			\hline\hline \noalign{\smallskip}
			Ion & $\lambda_{\rm obs}$ & $I_{\rm obs}$ & $T^{\rm (1)}_{\rm eff}$ & $T^{\rm (2)}_{\rm eff}$ & $R^{\rm (1)}$ & $R^{\rm (2)}$ \\
			\noalign{\smallskip}\hline\noalign{\smallskip}
			
 \ion{C}{i} & 1560.70 & $160.0^d$ &  4.06 &  4.05 & 1.49 & 1.33 \\
 \ion{C}{i} & 1560.31 & $130.0^d$ &  4.06 &  4.05 & 0.60 & 0.55 \\
 \ion{Si}{ii} & 1264.74 & $149.0^c$ &  4.25 &  4.24 & 0.61 & 0.88 \\
 \ion{S}{ii} & 1253.80 & $15.9^c$ &  4.27 &  4.26 & 1.00 & 1.45 \\
 \ion{C}{ii} & 1335.70 & $1205.0^c$ &  4.43 &  4.28 & 0.62 & 0.83 \\
 \ion{C}{ii} & 1036.30 & $35.9^b$ &  4.54 &  4.37 & 1.14 & 0.98 \\
 \ion{N}{ii} & 1085.70 & $36.7^b$ &  4.56 &  4.52 & 0.74 & 0.89 \\
 \ion{Si}{iii} & 1206.50 & $694.6^c$ &  4.66 &  4.63 & 0.63 & 0.62 \\
 \ion{C}{iii} &  977.00 & $702.0^b$ &  4.78 &  4.69 & 0.80 & 0.98 \\
 \ion{C}{iii} & 1175.74 & $104.0^b$ &  4.78 &  4.70 & 0.88 & 1.03 \\
 \ion{N}{iii} &  991.60 & $47.2^c$ &  4.82 &  4.84 & 0.86 & 0.75 \\
 \ion{O}{iii} &  703.85 & $43.5^a$ &  4.92 &  4.85 & 1.12 & 1.16 \\
 \ion{N}{iii} &  685.70 & $23.7^c$ &  4.88 &  4.91 & 1.19 & 1.09 \\
 \ion{S}{iv} &  661.40 & $6.9^c$ &  5.05 &  5.06 & 0.77 & 0.83 \\
 \ion{O}{iv} &  554.10 & $40.1^a$ &  5.25 &  5.19 & 1.09 & 1.03 \\
 \ion{O}{iv} &  554.55 & $111.7^a$ &  5.25 &  5.19 & 1.00 & 0.94 \\
 \ion{Ne}{iv} &  543.91 & $8.3^a$ &  5.28 &  5.27 & 0.76 & 0.80 \\
 \ion{Ne}{iv} &  542.10 & $4.6^a$ &  5.28 &  5.27 & 0.91 & 0.97 \\
 \ion{O}{v} &  629.78 & $338.5^a$ &  5.38 &  5.34 & 1.16 & 1.22 \\
 \ion{Ne}{v} &  572.31 & $8.8^a$ &  5.47 &  5.47 & 0.98 & 0.97 \\
 \ion{Ne}{vi} &  562.81 & $15.6^a$ &  5.67 &  5.67 & 1.09 & 1.08 \\
 \ion{Mg}{vii} &  435.20 & $28.3^c$ &  5.83 &  5.83 & 0.94 & 0.93 \\
 \ion{Mg}{vii} &  367.67 & $21.0^a$ &  5.88 &  5.88 & 1.00 & 1.00 \\
 \ion{Mg}{viii} &  436.70 & $42.7^c$ &  5.95 &  5.95 & 0.99 & 1.00 \\
 \ion{Mg}{viii} &  315.02 & $71.8^a$ &  5.96 &  5.96 & 0.81 & 0.82 \\
 \ion{Si}{viii} &  319.83 & $69.2^a$ &  6.01 &  6.01 & 0.89 & 0.89 \\
 \ion{Fe}{xi} &  352.67 & $30.5^a$ &  6.13 &  6.13 & 1.06 & 1.06 \\
 \ion{Si}{x} &  356.03 & $24.2^a$ &  6.14 &  6.14 & 0.93 & 0.93 \\
 \ion{Si}{x} &  347.40 & $44.7^a$ &  6.15 &  6.15 & 0.98 & 0.98 \\
 \ion{Fe}{xii} &  364.45 & $34.1^a$ &  6.19 &  6.19 & 1.06 & 1.06 \\
 \ion{Si}{xi} &  303.34 & $124.6^a$ &  6.22 &  6.22 & 1.17 & 1.17 \\
 \ion{Fe}{xvi} &  360.60 & $44.8^c$ &  6.38 &  6.38 & 0.76 & 0.76 \\			
			\noalign{\smallskip}\hline
			\noalign{\smallskip}
	\multicolumn{7}{p{0.9\columnwidth}}{\textbf{Notes.} Ion - principal ion emitting at observed wavelength $\lambda_{\rm obs}$ (\AA); $I_{\rm obs}$ - the measured radiance (ergs cm$^{-2}$ s$^{-1}$ sr$^{-1}$) from: a) Warren, b) Wilhelm et al., c) Vernazza and Reeves, and d) Sandlin et al.; $T_{\rm eff}$ - the effective temperature for each line (log values, in K), and $R$ - the ratio between the predicted and observed intensities using: (1) \textsc{Chianti} coronal approximation ion fractions, and (2) density dependent ion fractions for oxygen and carbon and \textsc{Chianti} for all other elements.}
		\end{tabular}
	\normalsize
	\label{tab:linesdem}
\end{table}

\section{Data availability}

New data produced in this work, direct ionisation and excitation--auto-ionisation by electron impact and oxygen ion fractions from the ionisation equilibrium, are available at CDS via anonymous ftp to cdsarc.u-strasbg.fr (130.79.128.5) or via http://cdsarc.u-strasbg.fr/viz-bin/qcat?J/MNRAS.

\bsp	
\label{lastpage}
\end{document}